\documentclass[english,12pt]{article}
\usepackage{physics} %to get the \order expansion right
\usepackage[symbol]{footmisc} %to get the footnotes right
\usepackage{hyperref} %to get the links in the text
\usepackage{cancel} %to cross out text

\hypersetup
{colorlinks   = true, 
 urlcolor     = black,
 linkcolor    = black, 
 citecolor    = black}
\usepackage{array}
 \usepackage[normalem]{ulem}
\usepackage{graphicx}
\usepackage{amssymb}
\usepackage[english]{babel}
\usepackage{amsmath}
\usepackage{multirow}
\usepackage{prettyref}
\usepackage{babel}
\usepackage{units}
\usepackage[latin1]{inputenc}
\usepackage{amsfonts}
\usepackage{amssymb}
\usepackage{babel}
\usepackage{color}
\usepackage[dvipsnames]{xcolor} 
\usepackage{cite}
\def\@fmsl@sh#1#2#3{\m@th\ooalign{$\hfil#1\mkern#2/\hfil$\crcr$#1#3$}}
 \def\eq#1\en{\begin{equation}#1\end{equation}}
\def\s[#1,#2]{[#1\stackrel{\star}{,}#2]}
\def\sx[#1,#2]{[#1\stackrel{\star_{x}}{,}#2]}

\newcommand{\logbox}{{\ln\left(\frac{\Box}{\mu^2}\right)}}

\newcommand{\GN}{{G_{\rm N}}}

\def\gsim{\mathrel{\rlap{\lower4pt\hbox{\hskip1pt$\sim$}}
		\raise1pt\hbox{$>$}}}       %greater than or approx. symbol

\newcommand{\nc}{\newcommand}
\nc{\beq}{\begin{equation}}
\nc{\eeq}{\end{equation}}
\nc{\beqa}{\begin{eqnarray}}
\nc{\eeqa}{\end{eqnarray}}

\def\bc{\begin{center}}
\def\ec{\end{center}}

\def\to{\rightarrow}

\def\gsim{\mathrel{\mathpalette\atversim>}}

\def\bc{\begin{center}}
\def\ec{\end{center}}

\def\gsim{\mathrel{\rlap{\lower4pt\hbox{\hskip1pt$\sim$}}

    \raise1pt\hbox{$>$}}}       %greater than or approx. symbol

\def\gsim{\mathrel{\rlap{\lower4pt\hbox{\hskip1pt$\sim$}}
    \raise1pt\hbox{$>$}}}       %greater than or approx. symbol

\begin{document}
\makeatletter
\def\fmslash{\@ifnextchar[{\fmsl@sh}{\fmsl@sh[0mu]}}
\def\fmsl@sh[#1]#2{%
  \mathchoice
    {\@fmsl@sh\displaystyle{#1}{#2}}%
    {\@fmsl@sh\textstyle{#1}{#2}}%
    {\@fmsl@sh\scriptstyle{#1}{#2}}%
    {\@fmsl@sh\scriptscriptstyle{#1}{#2}}}
\def\@fmsl@sh#1#2#3{\m@th\ooalign{$\hfil#1\mkern#2/\hfil$\crcr$#1#3$}}
\makeatother
%\baselineskip 24pt

%%%%%%%%%%%%%%%%%%%%%%%%%%%%%%%%%%%%%%%%%%%%%%%%%%%%%%%%%%%%%%%%%
%%%
%%%                      TITLE PAGE
%%%
%%%%%%%%%%%%%%%%%%%%%%%%%%%%%%%%%%%%%%%%%%%%%%%%%%%%%%%%%%%%%%%%%
\thispagestyle{empty}
\begin{titlepage}
\boldmath
\begin{center}
  \Large {\bf Black Hole Solutions in Quantum Gravity with Vilkovisky-DeWitt Effective Action}
    \end{center}
\unboldmath
\vspace{0.2cm}
\begin{center}
{\large Xavier~Calmet}\footnote[1]{x.calmet@sussex.ac.uk},
{\large Andrea~Giusti}\footnote[2]{a.giusti@sussex.ac.uk}
 {\large and}  
{\large Marco~Sebastianutti}\footnote[3]{m.sebastianutti@sussex.ac.uk}
 \end{center}
\begin{center}
{\sl Department of Physics and Astronomy, 
University of Sussex, Brighton, BN1 9QH, United Kingdom
}
\end{center}
\vspace{5cm}
\begin{abstract}
\noindent
We study new black hole solutions in quantum gravity. We use the Vilkovisky-DeWitt unique effective action to obtain quantum gravitational corrections to Einstein's equations. In full analogy to previous work done for quadratic gravity, we find new black hole like solutions. We show that these new solutions exist close to the horizon and in the far-field limit.
\end{abstract}  
\end{titlepage}

%\pacs{}

%%%%%%%%%%%%%%%%%%%%%%%%%%%%%%%%%%%%%%%%%%%%%%%%%%%%%%%%%%%%%%%%
%%%
%%%                     INTRODUCTION
%%%
%%%%%%%%%%%%%%%%%%%%%%%%%%%%%%%%%%%%%%%%%%%%%%%%%%%%%%%%%%%%%%%%

\newpage
\renewcommand{\thefootnote}{\arabic{footnote}}

In this letter we study new black hole solutions in quantum gravity. Following \cite{Lu:2015cqa,Lu:2015psa}, we assume the existence of a black hole horizon at $r=r_0$ and focus on the near-horizon behavior of these solutions. We use the Vilkovisky-DeWitt unique effective action \cite{Barvinsky:1983vpp,Barvinsky:1985an,Barvinsky:1987uw,Barvinsky:1990up,Buchbinder:1992rb,Calmet:2018elv} to derive quantum gravitational corrections to Einstein's equations and work to second order in curvature. These corrections are universal and model independent and apply to any ultra-violet complete theory of quantum gravity that admits General Relativity as a low energy limit. The corrections to Einstein's theory can be classified as local terms which come from parts of the effective action of the form $R^2$, $R_{\mu\nu}R^{\mu\nu}$ or $R_{\mu\nu\alpha\beta}R^{\mu\nu\alpha\beta}$ and non-local (in the sense that they involve derivative operators) terms 
$R\log\Box R$, $R_{\mu\nu}\log\Box R^{\mu\nu}$ or $R_{\mu\nu\alpha\beta}\log\Box R^{\mu\nu\alpha\beta}$. The Wilson coefficients on the local part are not calculable from first principles unless the effective action is matched to an ultra-violet complete theory of quantum gravity  \cite{Calmet:2024neu}. On the other hand, the Wilson coefficients of the non-local part are universal and can be calculated without specifying the ultra-violet complete theory. 

Finding new solutions in quantum gravity is a very challenging task. Fortunately, Stelle and his collaborators \cite{Lu:2015cqa,Lu:2015psa} have studied such solutions in his quadratic gravity model \cite{Stelle:1977ry}. His model matches the local part of the unique effective action and his results can thus be extended to the case under consideration here. Generalizing the results of \cite{Lu:2015cqa} to the unique effective action truncated at second order in curvature is straightforward because the local part and non-local part of the action are related by renormalization group invariance and we can thus easily deduce the new black hole-like solutions in quantum gravity using the results of \cite{Lu:2015cqa}  and by imposing renormalization group invariance of the new solutions. 

Our starting point is the unique effective action at second order in the curvature expansion \cite{Barvinsky:1983vpp,Barvinsky:1985an,Barvinsky:1987uw,Barvinsky:1990up,Buchbinder:1992rb,Calmet:2018elv}
\begin{equation}\label{eq:UEA}
    \Gamma_{\rm QG} = \Gamma_{\rm L} + \Gamma_{\rm NL} +\Gamma_{\rm matter}
\end{equation}
with a local part
\begin{align}
    \Gamma_{\rm L} 
    &= 
    \int d^4x \, \sqrt{|g|} \left[ \frac{M_P^2 }{2}
    \big(R - 2\Lambda\big)
    + c_1(\mu) \, R^2 
    + c_2(\mu) \, R_{\mu\nu} R^{\mu\nu} 
    \right.\nonumber\\
    &\qquad \qquad \qquad \qquad
    + c_3(\mu) \, R_{\mu\nu\rho\sigma} R^{\mu\nu\rho\sigma} 
    + c_4(\mu) \, \Box R
    + O(M_P^{-2}) \Big],
\end{align}
a non-local part 
\begin{align}
    \Gamma_{\rm NL} &= 
    - \int d^4x \, \sqrt{|g|} \left[ 
   \alpha \, R \ln \left(\frac{\Box}{\mu^2} \right) R
    +\beta \, R_{\mu\nu} \ln \left( \frac{\Box}{\mu^2} \right) R^{\mu\nu}
    \right.\nonumber\\
    &\qquad \qquad \qquad \qquad \left.
    +\gamma \, R_{\mu\nu\rho\sigma} \ln \left( \frac{\Box}{\mu^2} \right) R^{\mu\nu\rho\sigma}
    + O(M_P^{-2}) \right],
\end{align}
and with the matter sector enclosed in $\Gamma_{\rm matter}$. In this work, $M_P=\sqrt{\hbar \, c/(8 \,\pi \, G_N)}= 2.4\times10^{18}\,{\rm GeV}$ denotes the reduced Planck mass, $\Lambda$ the cosmological constant (which we disregard from now on), $c_i$ and $\alpha$, $\beta$ and $\gamma$ are Wilson coefficients.\footnote{Notice that we ignore the $c_4(\mu)$ term in the action as it is a total derivative.}

\begin{table}
\center
\begin{tabular}{| c | c | c | c |}
\hline
 & $ \alpha $ & $\beta$ & $\gamma$  \\
 \hline
 \text{Scalar} & $ 5(6\xi-1)^2$ & $-2 $ & $2$     \\
 \hline
 \text{Fermion} & $-5$ & $8$ & $7 $ \\
 \hline
 \text{Vector} & $-50$ & $176$ & $-26$ \\
 \hline
 \text{Graviton} & $250$ & $-244$ & $424$\\
 \hline
\end{tabular}
\caption{Non-local Wilson coefficients for different fields.
All numbers should be divided by $11520\pi^2$. Here, $\xi$ denotes the value of the non-minimal coupling for a scalar theory.}
\label{coeff1}
\end{table}

The local Wilson coefficients $c_i(\mu)$ depend on the running energy scale $\mu$. They are only calculable, if the effective action is matched to a UV complete theory of quantum gravity, see~\cite{Calmet:2024neu}. They are thus model-dependent. On the other hand, the values of the non-local Wilson coefficients $\alpha$, $\beta$ and $\gamma$ can be calculated from first principles without specifying a UV completion. They are model-independent and they depend only on the nature of the degrees of freedom that have been integrated out, see Table~\ref{coeff1}. The renormalization group equations for the local Wilson coefficients are given by 
\begin{align}
    c_1(\mu)&=c_1(\mu_*)-\alpha\ln{\left(\frac{\mu^2}{\mu^2_*}\right)},\\
    c_2(\mu)&=c_2(\mu_*)-\beta\ln{\left(\frac{\mu^2}{\mu^2_*}\right)},\\
    c_3(\mu)&=c_3(\mu_*)-\gamma\ln{\left(\frac{\mu^2}{\mu^2_*}\right)}.
\end{align}

Using the local and non-local Gauss-Bonnet identities one can absorb the Riemann tensor terms in~\eqref{eq:UEA} into the Ricci scalar and Ricci tensor ones~\cite{Calmet:2018elv}. Using these relations, we can rewrite the local and non-local effective action as
\begin{eqnarray}
    \Gamma_{\rm L}&=&\int d^4x\sqrt{\abs{g}}\left[\frac{R}{16\pi\GN}+\bar{c}_1(\mu)R^{2}+\bar{c}_2(\mu)R_{\mu\nu}R^{\mu\nu}\right],\\
    \Gamma_{\rm NL}&=&-\int d^4x\sqrt{\abs{g}}\left[\bar{\alpha} R\ln{\left( \frac{\Box}{\mu^2}\right)}R+\bar{\beta} R_{\mu\nu}\ln{\left( \frac{\Box}{\mu^2}\right)}R^{\mu\nu}\right],
\end{eqnarray}
where $\bar{c}_1(\mu)=c_1(\mu)-c_3(\mu)$, $\bar{c}_2(\mu)=c_2(\mu)+4c_3(\mu)$ and $\bar{\alpha}=\alpha-\gamma$, $\bar{\beta}=\beta+4\gamma$. The quantum corrected Einstein equations at second order in curvature are given by
\begin{equation}\label{eq:EOMS}
    E_{\mu\nu}\equiv R_{\mu\nu}-\frac{1}{2}Rg_{\mu\nu}+16\pi\GN\left(H_{\mu\nu}^{\rm L}+ H_{\mu\nu}^{\rm NL}\right)=8\pi\GN T_{\mu\nu},
\end{equation}
where 
\begin{align}
    &H_{\mu\nu}^{\rm L}=\bar{c}_1(\mu)\left(2R_{\mu\nu}R-\frac{1}{2}g_{\mu\nu}R^2+2g_{\mu\nu}\nabla^2R-\nabla_\mu\nabla_\nu R-\nabla_\nu\nabla_\mu R\right)\notag\\
    &+\bar{c}_2(\mu)\left(-\frac{1}{2}g_{\mu\nu}R_{\rho\sigma}R^{\rho\sigma}+2R_{\mu}^{\,\,\,\rho}R_{\nu\rho}+\nabla^2R_{\mu\nu}-\nabla_\rho\nabla_\mu R_{\nu}^{\,\,\,\rho}-\nabla_\rho\nabla_\nu R_{\mu}^{\,\,\,\rho}+g_{\mu\nu}\nabla_{\sigma}\nabla_{\rho}R^{\rho\sigma}\right),
\end{align}
and
\begin{align}
    &H_{\mu\nu}^{\rm NL}=\,-\bar{\alpha}\bigg(2R_{\mu\nu}-\frac{1}{2}g_{\mu\nu}R+2g_{\mu\nu}\nabla^2-\nabla_\mu\nabla_\nu-\nabla_\nu\nabla_\mu\bigg)\ln{\left( \frac{\Box}{\mu^2}\right)}R\notag\\
    &-\bar{\beta}\left(-\frac{1}{2}g_{\mu\nu}R^{\rho\sigma}+2\delta^{\sigma}_{\,\nu}R_{\mu}^{\,\,\rho}+\delta^{\rho}_{\,\mu}\delta^{\sigma}_{\,\nu}\nabla^2-\delta^{\sigma}_{\,\nu}\nabla^{\rho}\nabla_{\mu}-\delta^{\sigma}_{\,\mu}\nabla^{\rho}\nabla_{\nu}+g_{\mu\nu}\nabla^{\sigma}\nabla^{\rho}\right)\ln{\left( \frac{\Box}{\mu^2}\right)}R_{\rho\sigma}.\label{eq:HNL}
\end{align}

Considering a static and spherically symmetric metric of the form
\begin{equation}
    ds^2=h(r)dt^2-\frac{1}{f(r)}dr^2-r^2d\theta^2-r^2\sin^2{\theta}\,d\phi^2,
\end{equation}
and given the generalized Bianchi identity $\nabla^{\nu}E_{\mu\nu}=0$, it is easy to see that there are only two independent field equations, which for simplicity we choose to be $\left\{E_{tt}(r)=0,E_{rr}(r)=0\right\}$.

It has been show that at second order in curvature \cite{Calmet:2017qqa,Calmet:2018elv}, Schwarzschild is still a solution in quantum gravity. Corrections to Schwarzschild metric appear at third order in curvature \cite{Calmet:2021lny}. Here, we shall work at second order in curvature and we can thus follow the procedure highlighted in  \cite{Lu:2015psa,Lu:2015cqa}. We assume that the new solutions have an horizon, i.e. that they are black hole like and denote this horizon by $r_0$. We look for solutions that have small deviations from the Schwarzschild one close to $r_0$.
In the case of Schwarzschild, one has $f(r)=1-2 G M/r$. 
We emphasize that these new solutions are not corrections to  the classical Schwarzschild one, but rather new static vacuum solutions in full quantum gravity. Such solutions could exist as Birkhoff's theorem does not hold in quantum gravity  \cite{Calmet:2017qqa,Calmet:2018elv}. As shown in\cite{Nelson:2010ig} any static black hole solution of local part of Eq. (\ref{eq:EOMS}) must have vanishing Ricci scalar. Using \cite{Calmet:2018elv}, it is straightforward to show that this applies to the full field equations (\ref{eq:EOMS}) including the non-local part of the theory. 

%As explained in  \cite{Lu:2015psa,Lu:2015cqa} this enables us to drop terms in $R^2$  as they do not contribute to the field equations at which simplifies the calculations remarkably, however the authors of \cite{Lu:2015psa,Lu:2015cqa} also show that this does not imply that the Ricci tensor vanishes. We can thus consider the simplified action

We can rewrite the unique effective action in terms of the Weyl tensor
\begin{eqnarray}
S &=&\int d^4x \sqrt{|g|}\left(\frac{M_P^2}{2}R+ \tilde c_1(\mu) R^2 + \tilde c_2(\mu)
C_{\mu\nu\alpha\beta}  C^{\mu\nu\alpha\beta} \right.  \\  \nonumber
&& \hspace{4cm} \left. + \tilde \alpha R\ln{\left( \frac{\Box}{\mu^2}\right)}R+
\tilde \beta
C_{\mu\nu\alpha\beta} \ln{\left( \frac{\Box}{\mu^2}\right)} C^{\mu\nu\alpha\beta}
 \right),
\end{eqnarray}
where $\tilde c_1(\mu)=c_1(\mu)+(c_2(\mu)+c_3(\mu))/3$,
$\tilde c_2(\mu)=c_2(\mu)/2+2c_3(\mu)$, $\tilde \alpha=\alpha+(\beta+\gamma)/3$ and $\tilde \beta=\beta/2+2\gamma$. The contributions from the $R^2$ and $R\log \Box R$ terms to to the field equations are proportional to $R$ and thus vanishing for black hole-like solutions and we can thus focus on the Weyl action setting $\tilde c_1=0$ and $\tilde \alpha=0$.

As in \cite{Lu:2015psa,Lu:2015cqa} for the specific case of quadratic gravity, we now suppose that for any theory of quantum gravity that admits General Relativity as a low energy effective action,  there exists a black-hole horizon at some
radius $r = r_0>0$, at which the metric functions $f(r)$ and
$h(r)$ vanish, and we then obtain near-horizon Taylor expansions for the metric functions of the type:
\begin{eqnarray}
h(r)&=& k  \left ( ( r-r_0) + h_2(r-r_0)^2 + h_3(r-r_0)^3 \right)+ {\cal O}((r-r_0)^4) \\
f(r)&=& f_1( r-r_0) + f_2(r-r_0)^2 + f_3(r-r_0)^3+ {\cal O}((r-r_0)^4). 
\end{eqnarray}
Note that we are focusing on black holes with masses much larger than the Planck mass, i.e. classical black holes. Close to the horizon of these black holes, curvature is weak and we can thus safely trust our curvature expansion and truncate the effective action at second order. The second-order curvature expansion is reliable for macroscopic black holes, where the curvature near the horizon is weak for black holes of masses much larger than the Planck mass (see e.g. \cite{Calmet:2021lny}).

It is straightforward to calculate the Ricci scalar and the Ricci tensor in terms of $f(r)$ and $h(r)$. The Ricci scalar and tensor can then be inserted in \eqref{eq:EOMS}. While the local part of the field equations is identical to that of quadratic gravity studied in \cite{Lu:2015psa,Lu:2015cqa}, the non-local part represents a considerable challenge as we need to calculate
$\logbox C_{\mu\nu\alpha\beta}$. This is not straightforward because the components of the  Weyl tensor are given in terms of complicated fractions involving the coefficients $f_i$ and $h_i$. We could in principle perform a partial fraction expansion before folding the individual terms with the distribution corresponding to the $\log$ term. However, there is a much more straightforward manner of obtaining the result using renormalization group invariance: we can focus on deriving the local part of the field equations obtaining a metric which has a renormalization  scale dependence and then deduce the contribution of the non-local part of the field equations in~\eqref{eq:EOMS} by imposing renormalization group invariance of the metric. The $\mu$ dependence in the local part of the field equations must be canceled by the $\mu$ contribution coming from the non-local part.

The renormalization group invariant equations of motion are obtained by replacing $\bar{c}_i(\mu)$  with
\begin{align}
    \bar{c}_1(\mu)&\to\bar{c}_1(\mu)+2\bar{\alpha}\ln{\mu},\notag\\
    \bar{c}_2(\mu)&\to\bar{c}_2(\mu)+2\bar{\beta}\ln{\mu},
    \notag\\
    \bar{c}_3(\mu)&\to\bar{c}_3(\mu)+2\bar{\gamma}\ln{\mu},\label{eq:cbars}
\end{align}
in the local part of the field equations.  

We can now determine the functions $f_i$ and $h_i$. The constant $k$ is arbitrary and linked to the freedom to rescale the time coordinate. The leading order terms are given by
\begin{eqnarray}
h_2&=& \frac{1-2 f_1 r_0}{f_1 r_0^2} + \frac{1- f_1 r_0}{8 (\tilde c_2(\mu)+2 \tilde \beta \ln{\mu}) f_1^2 r_0 }\\
f_2&=& \frac{1-2 f_1 r_0}{r_0^2} - \frac{3(1- f_1 r_0)}{8 (\tilde c_2(\mu)+2 \tilde \beta \ln{\mu}) f_1 r_0 }
\end{eqnarray}
while higher order corrections can be obtained using the same procedure\footnote{Note that this expansion is strictly  speaking valid only for $r$ close $r_0$ and that away from $r_0$, we have $R={\cal O}(r-r_0)$. Furthermore, the effective stress-energy tensor is well-behaved at $r=r_0$ and its neighborhood.}. From there on the numerical analysis performed in   \cite{Lu:2015psa,Lu:2015cqa} applies fully. Note however that our results imply that the new black hole like metrics are renormalization group invariant and thus physical quantities can be obtained reliably from our metrics. Also our work is not specific to quadratic gravity. It applies to any ultraviolet complete theory of quantum gravity which has general relativity as a low energy limit. Our work implies that the non-Schwarzschild static black hole found numerically in \cite{Lu:2015psa,Lu:2015cqa} are present in any such ultraviolet complete theory. We stress again that these are new solutions in full quantum gravity and not corrections to  the classical Schwarzschild metric.

While the method developed in \cite{Lu:2015psa,Lu:2015cqa} based on a Frobenius ansatz for the metric functions $f(r)$ and $h(r)$  could be applied to obtain solutions near the origin, the effective theory will breakdown close to $r=0$ and the truncation of the quantum effective action at second order in curvature cannot be trusted. We thus cannot make any statement about the behavior of the new solutions at the origin and in particular it is not clear whether these new black hole solutions have a singularity at  $r=0$. One would expect that singularities are cured by quantum gravity, but this cannot be studied within the approach we are using. Note that the large $r$ limit had been considered in \cite{Calmet:2018egd,Calmet:2018hfb}. The metric receives corrections that correspond to the new massive spin-2 and spin-0 classical fields contained in the effective action.

In this letter, we discuss the existence of black hole-like solutions beyond the classical Schwarzschild solution in quantum gravity. Specifically, these solutions are solutions to the quantum corrected Einstein's equations obtained at second order in curvature using the Vilkovisky-DeWitt unique effective action. Asymptotically these solutions are indistinguishable  from Schwarzschild if the new classical degrees of freedom are heavy enough as their contributions are damped exponentially. However, they differ from the classical Schwarzschild close to the event horizon. Note that our results are another proof of the violation of Birkhoff's theorem in quantum gravity \cite{Calmet:2017qqa,Calmet:2021stu} as we have new solutions beyond Schwarzschild's one.

    \noindent {\it Acknowledgments:}
	The work of X.C. and A.G. is supported in part  by the Science and Technology Facilities Council (grants numbers ST/T006048/1 and ST/Y004418/1.). The work of M.S. is supported by a doctoral studentship of the Science and Technology Facilities Council (training grant No. ST/X508822/1, project ref. 2753640).
\\
	\smallskip 
	
	\noindent {\it Data Availability Statement:}
	This manuscript has no associated data. Data sharing not applicable to this article as no datasets were generated or analyzed during the current study.
	
	%\bigskip 

%%%%%%%%%%%%%%%%%%%%%%%%%%%%%%%%%%%%%%%%%%%%%%%%%%%%%%%%%%%%%%%%%
%%%
%%%                     BIBLIOGRAPHY
%%%
%%%%%%%%%%%%%%%%%%%%%%%%%%%%%%%%%%%%%%%%%%%%%%%%%%%%%%%%%%%%%%%%%

\bigskip{}

\end{document}